# Concepts for a Muon Accelerator Front-End

**Diktys Stratakis[a,*] Scott J. Berg[b], and David Neuffer[a]**

[a] *Fermi National Accelerator Laboratory,*
*Batavia, IL 60510 USA*
[b] *Brookhaven National Laboratory,*
*Upton, NY 11973 USA*
*E-mail*: diktys@fnal.gov

ABSTRACT: We present a muon capture front-end scheme for muon based applications. In this Front-End design, a proton bunch strikes a target and creates secondary pions that drift into a capture channel, decaying into muons. A series of rf cavities forms the resulting muon beams into a series of bunches of differerent energies, aligns the bunches to equal central energies, and initiates ionization cooling. We also discuss the design of a chicane system for the removal of unwanted secondary particles from the muon capture region and thus reduce activation of the machine. With the aid of numerical simulations we evaluate the performance of this Front-End scheme as well as study its sensitivity against key parameters such as the type of target, the number of rf cavities and the gas pressure of the channel.

KEYWORDS: Muon capture; Targetry; Pion decay.

[*]Corresponding author

# Contents

## 1. Introduction

Muon accelerators are being explored for a Neutrino Factory [1] and a Muon Collider [2]. Muons are commonly produced indirectly through pion decay by the interaction of a high-energy (multi-GeV), high-power (1-4 MW) proton beam with a target [4]. The muon yield is fractionally small, with large angle and energy dispersion, so that efficient collection is necessary in all dimensions of phase-space [5-7].

There are two main requirements on the Front-End system, which is located between the target and the acceleration systems: First, given the short lifetime of muons – 2.2 µs in the rest frame – the Front-End system has to collect the pions and form a beam from their daughter muons as quickly as possible. Second, the front-end has to manipulate the transverse and longitudinal phase-space of the beam so that it matches the acceptance criteria of the downstream accelerators.

In this paper, we present a muon capture scheme [8] that is suitable for muon based accelerators. In that scheme, the pions (and resulting muons) are initially produced within a short bunch length and a broad energy spread. The $\pi$'s drift from the production target, lengthening into a long bunch with a high-energy "head" and a low-energy "tail", while decaying into µ's. The beam is then transported through a "Buncher" that forms the beam into a string of bunches, and then an "rf Rotator" section that aligns the bunches to (nearly) equal central energies. The µ's are then cooled in a "Cooler" with rf cavities and absorbers. An important feature of the present scheme is that it captures and cools both signs ($\mu^+$ and $\mu^-$) simultaneously, at roughly equal efficiencies.

With the aid of numerical simulations, we show that our proposed muon collection scheme can capture muons with a notable rate of 0.12 per incident 8 GeV proton. We systematically analyse the sensitivity in performance of the channel against key parameters such as the number of cavities and the gas pressure of the transfer line. We also examine the performance for two different target scenarios: one based on a liquid mercury and another based on a solid carbon target.

In this paper, we also examine the effect of undesirable secondary particles exiting the target region and passing through the subsequent muon capture systems. Hadronic pollutants in the beam tend to cause activation of accelerator components, preventing hands-on maintenance of the machine. This would lead to additional construction and operation costs and is highly undesirable. Leptonic pollutants in the beam cause less activation of accelerator components but are still undesirable due to the increased heat load that may be placed on superconducting components. Our proposed system comprises a solenoidal chicane that filters high momentum particles (> 1 GeV/c), followed by a absorber that reduces the energy of all particles, resulting in the rejection of low energy species that pass through the chicane. We detail the design and optimization of this system and its integration with the rest of the muon capture scheme.

## 2. Muon capture components

The main components of the muon capture scheme are illustrated in Fig. 1. In the subsections below we review those components:

### 2.1 Target

In the present baseline configuration, the input beam was taken to be composed of a 8 GeV proton beam with a 2 ns time spread incident on a liquid mercury target in the bore of a 20 T solenoid [10]. The proton beam and mercury jet are tilted with respect to the solenoidal field magnetic axis, so that non-interacting particles impinge on the mercury jet collection pool which acts as the proton dump [11]. The interaction of the beam with the target was modeled using MARS [12]. We then propagate the produced particles using ICOOL [13]. After the target, the field tapers from 20 T to 2.0 T, collecting both positive and negative particle species (see Fig. 1). Then, the field continues at 2 T for another 15.25 m. The radius in the decay channel as well as in all subsequent sections is assumed to be 0.25 m.

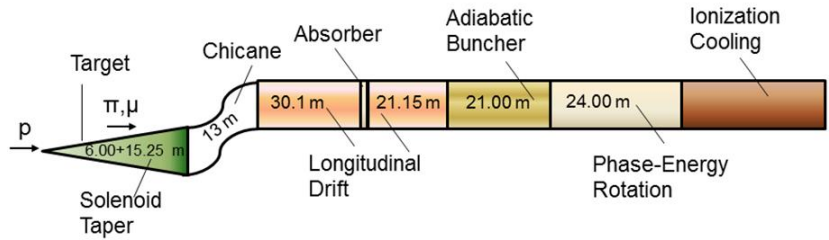

**Figure 1**: Illustration of a muon collection system for muon accelerators.

A major improvement in our present design is that the magnetic field drops [14] from its peak value down to 2 T over 6.0 m rather than the 14.75 m considered in previous studies [4]. Quantitatively, the effect of the solenoidal field profile on the lattice capture efficiency can be evaluated by calculating the muon yield, which is defined as the number of particles that fall within a reference acceptance, and this yield approximates the expected acceptance of the downstream accelerator. For the Neutrino Factory case, the transverse normalized acceptance is 30 mm and the normalized longitudinal acceptance is 150 mm [15]. The new profile increases the final muon yield by ~8% and this is a direct consequence of the fact that a short taper delivers a denser distribution of muons in longitudinal phase-space, which permits a more effective bunch formation in the buncher and phase-rotator sections further downstream [16].

## 2.2 Target material

In the present section, we compare the performance of the muon capture channel for two different target scenarios: one based on liquid mercury and another one based on a solid carbon [17]. In both cases the target is in a field that peaks near 20 T at the center of the beam-target crossing and tapers down to 2 T as described in the previous section. We produce distributions for the two different target materials and discuss differences in particle spectrum near the sources. We then propagate the distributions through our capture system and compare the full system performance for the two target types. Figure 2(a) displays the distribution of positive pions 2 m downstream from the beam-target crossing point for a mercury target with an 8 GeV incident proton beam and a carbon target with a 6.75 GeV incident proton beam. Figure 2(b) displays the same but for negative pions. The differences are much larger for negative than for positive pions. Clearly, pion production per unit of proton power in mercury is notably higher than in carbon for lower pion energies, but the production rapidly becomes closer above 250 MeV. This suggests that the particle capture system should be ideally optimized differently for a mercury than for a carbon target due to the lower-energy spectrum in mercury. Furthermore, the difference in the spectral shape between positive and negative pions in mercury indicates that the optimal capture parameters for positive and negative particles will be different, and some application dependent optimal compromise parameters should be chosen.

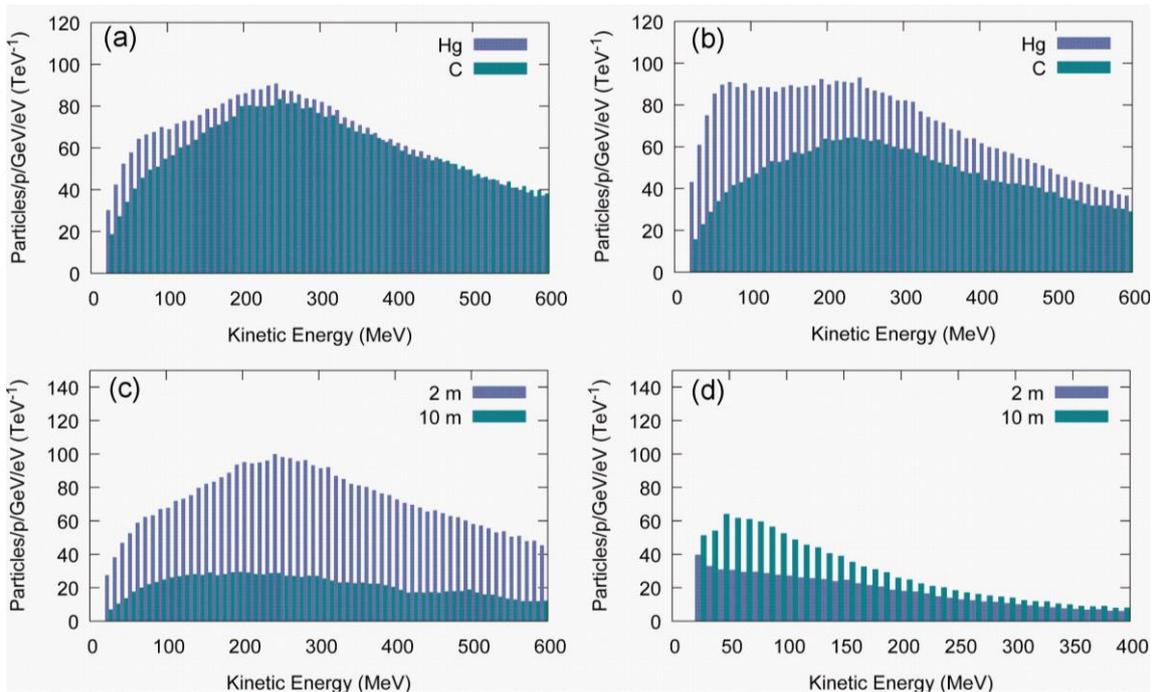

**Figure 2:** (a) Distribution of positive pions 2 m downstream from the beam-target crossing point for a mercury target with an 8 GeV incident proton beam and a carbon target with a 6.75 GeV incident proton beam; (b) Same for negative pions; (c) Positive pion spectra for a carbon target at two positions downstream from the beam-target crossing point, and (d) Positive muon spectra for a carbon target.

Figure 2(c) displays the positive pion spectra for a Carbon target at two different positions downstream from the beam target crossing point. Figure 2(d) shows the positive muon spectra for a Carbon target at two different positions from the beam-target crossing point. Clearly, a large number of pions are lost on the beam pipe, with a greater fractional loss for higher energy pions [see Fig. 2(c), and note that the pion loss far exceeds the muon gain in Fig. 2(d)]. Thus a higher solenoid field downstream (giving a smaller beam size) will lead to more particles transmitted and ultimately captured (consistent with results in [18]), but the capture system will need to be retuned for a higher energy range to make optimal use of the increased field. Finally, we also have inconclusive evidence that a small amount of absorber at large radius may increase the number of particles in the low energy portion of the spectrum, potentially leading to improved performance.

**2.3 Particle selection scheme**

The beam arising from the target is primarily made up of four constituent particles: protons, neutrons, pions and electrons. Additionally, some muons, kaons and other particles are present in the beamline. The muon capture system collects muons of both species in a momentum range of roughly 100–400 MeV/c, while all other particles are considered contaminants that contribute only to uncontrolled losses in the downstream systems. Protons are the main contaminant and their removal will be the focus of the present section. The peak at 8 GeV is primary protons that are lost in the target system, but there is a spectrum of secondary protons from the lowest energies to the highest that are captured and transported. Without collimation, this flux is lost on the capture channel apertures at kW/ m levels, much larger than the approximately 1 W/m desired to ensure "hands on" maintenance [19].

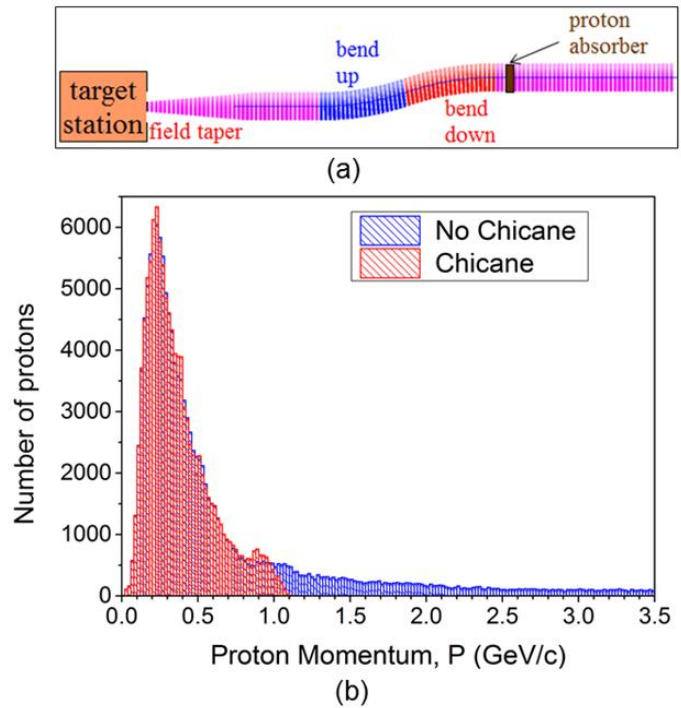

**Figure 3:** The capture region, including a chicane to suppress energetic protons and an absorber for soft protons; (b) distribution of protons 34.25 m downstream of the target with (red) and without (blue) a chicane. The chicane clearly removes high energy protons from the beam.

Rogers [9] proposed a solenoid bend chicane system to eliminate the high-energy protons and other unwanted secondaries and an absorber to deal with the remaining low energy particles. In the chicane system, high momentum particles are not strongly deflected by the bend

– 4 –

solenoid and are lost in or near the chicane and collimated on shielded walls. Lower momentum particles are strongly focused by the solenoid and follow the chicane with little orbit distortion. For our baseline system, the solenoidal chicane is placed at the end of the 15.25 m decay drift. The chicane bends out by $\theta = 15.0^0$ over $L = 6.5$ m and back by the same angle over 6.5 m more [Fig. 3(a)]. The chicane magnetic field is simulated in ICOOL using a toroidal model. Specifically, the magnetic field in each chicane section is modeled with a purely longitudinal field of $B_0 = (1 + \theta x/L)$, where $x$ is the horizontal coordinate, positive away from the center of curvature and $B_0$=2 T. There is a constant 2 T solenoidal field downstream from that point. Note that simulations using a field map generated by a set of coils produced the same results [20].

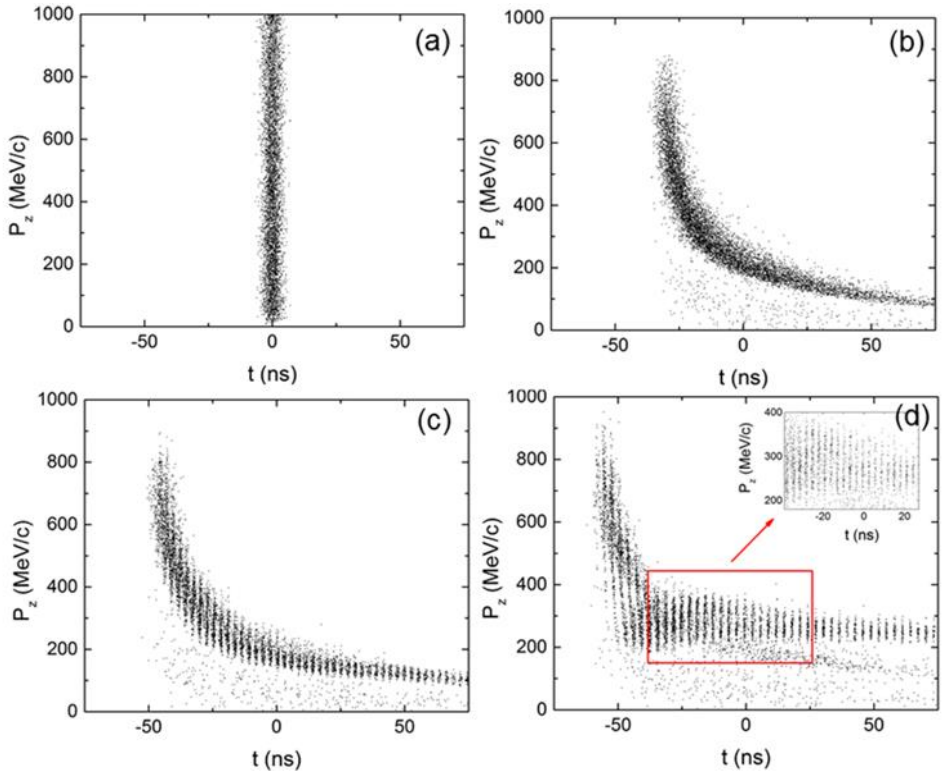

**Figure 4:** Evolution of the longitudinal phase-space along the drift, buncher, and phase-rotator within the front-end channel. (a) Pions and muons as produced at the target; (b) muons after the drift; (c) at the end of the buncher; and (d) at the end of the rotator. The present muon capture scheme produces 21 muon bunches (see red box) which are aligned into nearly equal energies.

The chicane can process particles of both signs and its action on the incident protons is displayed in Fig. 3(b). A glance at Fig. 3(b) indicates that high momentum particles (> 1 GeV/c) are lost in or near the chicane and collimated on shielded walls. The remaining particles pass through a 10 cm Beryllium absorber which removes almost all of the remaining low energy protons. The absorber would also stop pions before they decay to muons and was therefore moved 30.1 m downstream of the chicane, maximizing downstream muon transmission. An additional 21.15 m drift was added after the absorber to extend the beam distribution and obtain the energy-position correlation needed for bunching and phase-rotation. The chicane/absorber system was found to reduce the downstream energy deposition by more than an order of magnitude over that of a muon collection scheme without it [20].

Figure 4 shows the evolution of the muon beam longitudinal phase-space starting from the target [Fig. 4(a)]. As the beam drifts away, pions decay into muons, and the bunch lengthens, developing a "high energy head" and a "low energy tail". At the same time, the chicane chops away muons with momentum > ~1 GeV/c [Fig. 4(b)]. The separation of the particles that develops is given by:

$$\delta(ct_i) = L\left(\frac{1}{\beta_i} - \frac{1}{\beta_0}\right), \quad (2.1)$$

where $\delta(ct_i)$ indicates the time delay from a reference particle of speed $\beta_0$, $L$ is the distance from the production target and $\beta_i$ is the particle longitudinal speed.

## 2.4 Chicane optimization

We examine in this section the sensitivity in performance of the particle selection system by first scanning the geometric parameters of the chicane and looking for solutions with the best transmission that remove almost all the protons above a given energy (the "maximum proton kinetic energy", which we will henceforth denote with $K$). We use this to express the parameters of the chicane geometry in terms of $K$. We choose several of these optimal geometries and add a beryllium absorber downstream of the chicane, put it at two different positions, vary its thickness, and examine the muon transmission and the effectiveness of the system at removing protons.

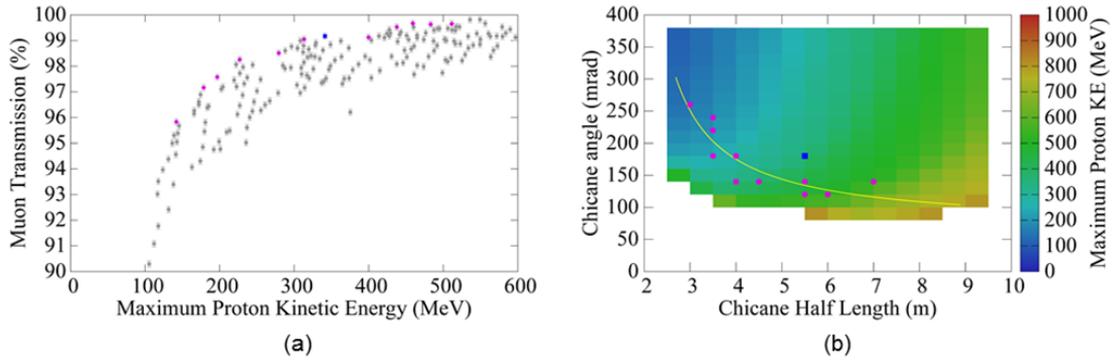

**Figure 5:** (a) Muon and pion transmission (as defined in text) and K for the chicane parameters we scanned. Magenta points were used to fit the chicane geometry parameters as a function of the ***K***. The square blue point was also originally selected, but was removed from the fit. Chicane parameters, and (b) ***K*** downstream of the chicane as a function of ***L*** and ***θ***. Points correspond to the colored points in Fig. 5(a). The curve shows the geometric parameters from Eq. 2.2.

We scan $\theta$ in 20 mrad steps and $L$ in 0.5 m steps. Our performance criteria are $K$ and the muon transmission, without an absorber, at a position 44.1 m downstream from the start of the chicane. The muon transmission is the number of the muons with kinetic energies between 80 and 260 MeV and pions with kinetic energies between 80 and 320 MeV, divided by the same quantity without a chicane. Figure 5 shows the results of that parameter scan. Chicanes with very different parameters can have similar $K$ but different transmissions. We chose some parameters which were on the high transmission edge of the points in Fig. 6(a). Those points are colored in the figure, and their θ and L are plotted in Fig. 6(b). We then fit those points to the functional forms:

$$L = L_0 + L_1 K \quad \theta = \theta_0 + \theta_1/K \quad (2.2)$$

The resulting parameters are $L_0$= 1.6 m, $L_1$= 9.1 m/GeV, $\theta_0$= 69 mrad and $\theta_1$=28 mrad GeV; we use these parameters whenever we evaluate Eq. 2.2.

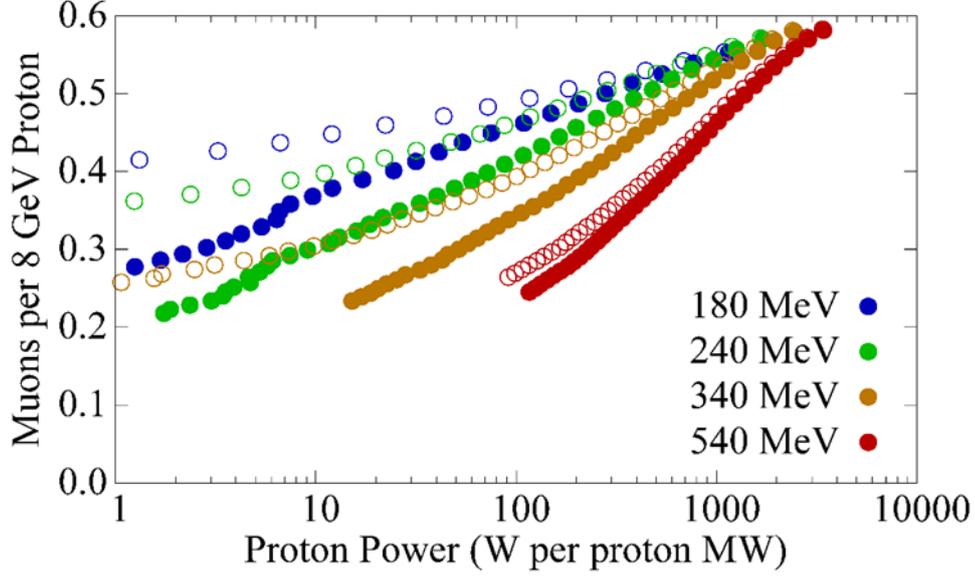

**Figure 6:** Parameters Each point shows, for a given chicane geometry and absorber position and thickness, the muons with kinetic energies in the range of 20 to 390 MeV and the proton power at a position 31 m from the beginning of the chicane. Each color is for a different chicane geometry, as defined by **K** (shown in the figure key) and Eq. 2.2. The absorber is positioned at the end of the chicane for filled circles, and with its upstream face 30 m from the beginning of the chicane for open circles. For each symbol, points for different absorber thicknesses in 1 cm, starting at 1 cm in the top right.

We next take parameters from Eq. 1 for four values of $K$ and create distributions at the end of the chicane. We then propagate the distributions downstream, passing the beam through a beryllium absorber. We vary the thickness of the beryllium absorber and try two locations for the absorber, one at the end of the chicane, the second with the front face of the absorber 30 m from the start of the chicane. 31 m downstream from the start of the chicane, we count the number of muons with kinetic energies between 20 and 390 MeV, and the energy of the protons. Figure 6 shows the results. We find better muon transmission for a given proton power downstream with chicane designs that have a lower $K$, and for the chicane positioned further downstream. For a given $K$ and absorber position, increased absorber thickness reduces proton power and muon transmission. The power allowed downstream will determine the optimal parameters.

### 2.5 Buncher and Phase-Rotation

The decay channel is followed by a 21 m buncher section in which the gradient of the rf system gradually increases and the beam is captured into a string of bunches with different energies. The rf frequency decreases along the length of the buncher, with the constraint that the phase difference between two reference particle momenta, $P_0$ and $P_N$, remains a fixed number of wavelengths as the beam propagates through it, i.e.,

$$N_B \lambda_{rf}(L) = L\left(\frac{1}{\beta_N} - \frac{1}{\beta_0}\right) \qquad (2.3)$$

where $\beta_0$ and $\beta_N$ are the velocities of the reference particles at momentum $P_0$ and $P_N$, respectively. Following this procedure, the reference particles and all intermediate bunch centers remain at zero crossings of the rf wave throughout the buncher. For the present design, $N_B$ is chosen to be 12, $P_0$=250 MeV/c and $P_N$=154 MeV/c with the intent of capturing particles within 50 to 400 MeV energy range. With these parameters, the rf frequency at the beginning of the buncher section is 490 MHz and at the end falls to 365 MHz. In the bunching system, 56 normal conducting pillbox-shaped rf cavities are employed, each having a different rf frequency and and rf gradient G that increases linearly with distance. This gradual increase of voltage enables a somewhat adiabatic capture of muons into separated bunches. The buncher consists of 28 cells, 0.75 m each, containing two 0.25 m long cavities. To keep the muon beam focused, a constant 2.0 T field is maintained through the section.

**Table 1:** rf frequencies of the buncher and phase rotator.

| Buncher frequency (MHz) | Buncher gradient (MV/m) | Rotator frequency (MHz) | Rotator gradient (MV/m) |
|---|---|---|---|
| 493.71 | 0.30 | 363.86 | 20 |
| 482.21 | 1.24 | 357.57 | 20 |
| 470.27 | 1.95 | 352.20 | 20 |
| 458.40 | 3.38 | 347.59 | 20 |
| 448.07 | 4.45 | 343.65 | 20 |
| 437.73 | 5.52 | 340.27 | 20 |
| 427.86 | 6.60 | 337.39 | 20 |
| 418.43 | 7.67 | 334.95 | 20 |
| 409.41 | 8.74 | 332.88 | 20 |
| 400.76 | 9.81 | 331.16 | 20 |
| 392.48 | 10.88 | 329.75 | 20 |
| 384.53 | 11.95 | 328.62 | 20 |
| 376.89 | 13.02 | 327.73 | 20 |
| 369.55 | 14.30 | 327.08 | 20 |
|  |  | 326.65 | 20 |
|  |  | 326.41 | 20 |

Once the beam leaves the buncher, it consists of a series of bunches at different energies [Fig. 4(c)]. The beam then is phase-rotated with a second string of cavities with decreasing frequencies, but with constant accelerating gradient. The frequencies are chosen so that the centers of the low- energy bunches increase in energy, while those of the high-energy bunches decrease. The algorithm used for setting this condition is to keep the first reference particle at fixed momentum while uniformly accelerating the second reference particle through the rotator section, so that it attains the first particle's energy at the end of the channel. With this condition, the bunches are aligned into nearly equal energies over the 24 m length of the rotator. The rf gradient is kept at 20 MV/m while the frequency of the rf decreases from 364 to 326.5 MHz. At the rotator exit the reference particles are at the same momentum ~245 MeV/c and the rf frequency is matched to 325 MHz. In a similar fashion to the buncher, the rotator cell is 0.75 m in length, contains two cavities (each has a length of 0.25 m). In total the rotator includes 32 cells with 64 rf cavities while the focusing field remains at 2.0 T throughout the rotator section.

The longitudinal phase-space distribution of the beam at the phase-rotator exit is shown in Fig. 4(d). For muon accelerator applications [21, 22], only 21 bunches (shown within the red box) are used for subsequent cooling and acceleration. Here it is important to note that because the focusing system consists of solenoids, the muon production, collection, bunching and phase rotation system produces bunches of both positive and negative muons at roughly equal intensities.

Initially we assumed a continuously decreasing frequency where each cavity has a different frequency leading to 120 rf frequencies in the buncher and phase-rotator combined. In a more realistic implementation, the cavities need to be paired into a smaller number of frequencies. Our simulations suggest that if the cavities are grouped into a pair of four, which corresponds to 30 discrete frequencies, the relative muon yield is reduced by ~9%. However, if the cavities are grouped into a pair of eight (15 discrete frequencies) the resulting muon drops by more than 20%. As a result, it is preferable to combine the cavities into groups of four and Table I has the required frequencies and rf cavity gradients.

## 2.6 Ionization cooling

While the magnetic field is constant in the decay channel, buncher and phase-rotator section, the field in the subsequent cooling channel is generated by alternating solenoid coils of ±3 T strength. We considered using 9 solenoid coils to match [18]. The magnet settings were optimized using a standard Nelder-Mead algorithm [23] with the objective to maximize the muon yield in the following ionization cooling section.

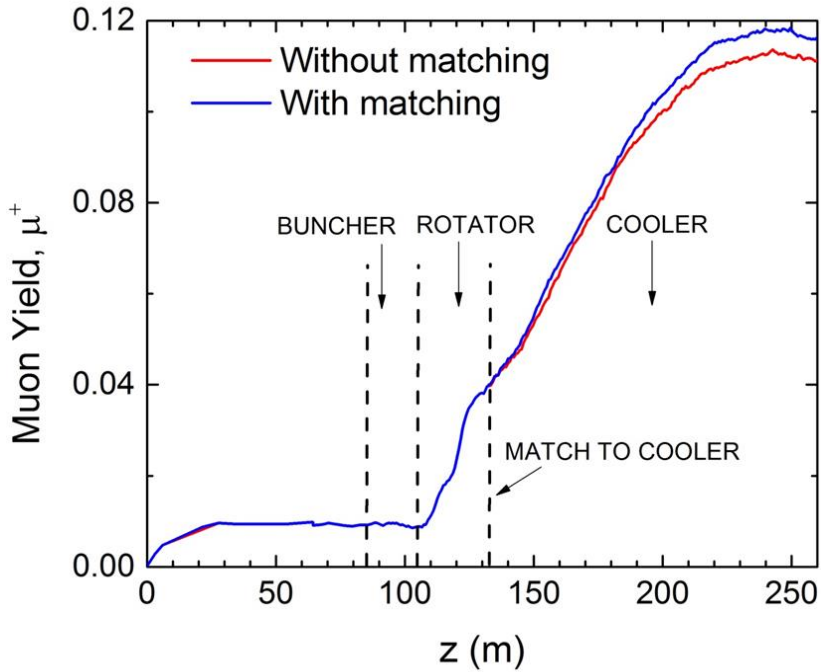

**Figure 7:** Muon yield along the muon capture channel with and without optimizing a matching section between the phase-rotator and cooler.

After the matching section, the muon beam enters a ionization cooling channel [24] consisting of rf cavities, absorbers, and alternating focusing solenoidal magnets. The channel consists of a sequence of ~70 identical 1.5 m long cells. Each cell contains 2 doublets of 0.25 m long 325 MHz pillbox rf cavities with 0.25 m spacing between the doublets and a 1.5 cm thick Lithium Hydride (LiH) discs at the ends of each doublet (four per cell). In addition, each cell contains two solenoid coils with opposite polarity, yielding an approximately sinusoidal variation of the magnetic field with a peak value of ~ 3 T on axis. The axial length of the coil is 15 cm, with an inner radius of 35 cm, an outer radius of 50 cm and a current density of 105.6 A/mm$^2$. Figure 7 shows the number of accepted muons (muon yield) along different parts of the channel before and after the match. It becomes evident that the inclusion of a matching section adds a 8% to the total gain. Acceptance is maximal at ~ 0.120 muons ($\mu^+$) per initial 8 GeV proton at z=240 m (~100 m of cooling).

## 2.7 Transport with high-pressure gas

The use of rf cavities filled with high pressure hydrogen gas was also examined as a potential solution for a muon capture front-end channel. The gas not only provides the necessary momentum loss as a cooling material but also protects the rf cavity from the multi-tesla magnetic field [25, 26]. We examine the performance for two different pressures: One at 34 atm and one at 100 atm, both at room temperature.

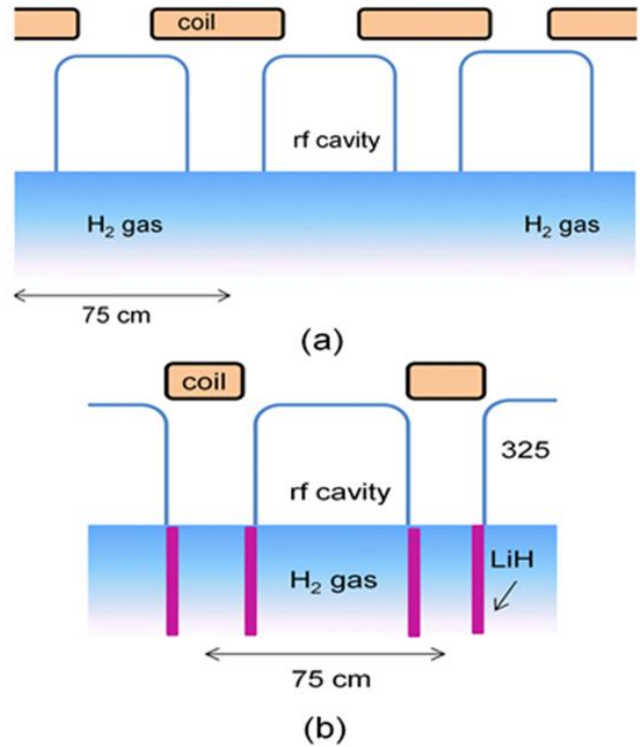

**Figure 8:** Gas filled scheme: (a) Buncher and phase rotator section, and (b) cooler section.

The principle for our proposed gas filled system is illustrated in Fig. 8 for the buncher/rotator system and the cooler. In a standard gas filled cooling channel the energy loss is distributed throughout the channel, rather than localized at discrete absorbers. One challenge is that such a system requires high pressure ( The density corresponding to ~160 atm at room temperature.). For the cooling system, we propose an alternative wherein we use a minimum value of gas sufficient enough to protect the rf while the majority of cooling is still done with LiH. We also propose using gas filled cavities in the phase rotator and buncher sections. The gas adds dE/ds energy loss to the muons passing through the buncher and rotator, as well as multiple scattering.

The energy loss from the gas is ~1.14 MV/m at 34 atm and 3.5 MV/m at 100 atm. To recover performance similar to or better than the gas-free case, this energy loss must be compensated by adding more rf gradient for acceleration and bunching.

The target and drift sections remain similar to the baseline case. Initially 34 atm (0.00285 gm/cc) is used for both buncher and rotator. If rf voltages are maintained at baseline, μ/p drops by 17%. If the rf gradients are increased to 20/25/28 for buncher/ rotator/ cooler, respectively, μ/p matches the baseline performance. Much of the loss occurs at the beginning of the buncher section where the rf gradient is relatively low and the gas based energy loss is relatively large. Reduction of the gas density in the buncher to 17 atm improves μ/p by 3% above the baseline.

An increase in density in the rotator to 100 atm (0.084gm/cc) would increase the amount of cooling in the rotator, and give added margin in avoiding breakdown. Increasing to 100 atm while keeping gradients at 20/25/28 MV/m, reduces the μ/p by 5% compared to the baseline. Increasing the rotator voltage to 28 MV/m, restores μ/p to the baseline performance. We conclude that the higher density can be accommodated with a moderate increase in rf gradients.

## 3. Summary

For muon accelerators, muons must be formed into short, intense bunches for maximal efficiency. Commonly, the muons are produced from bunches of protons that are focused onto a target to produce pions, that then decay into muons. The muons are produced within a very large phase space that must be captured and cooled to obtain high-luminosity parameters. In this study, we have described a technique for manipulating the longitudinal and transverse phase-space with the purpose of efficiently capturing and transporting a muon beam from the production target towards the accelerator chain. In that method, a set of properly tuned rf cavities captures the muon beams into strings of bunches and aligns them to nearly equal central energies, and a following set of rf cavities with absorbers cools them by a factor of three in transverse emittance. We found that our present muon scheme is capable of capturing simultaneously muons of both signs with a notable rate of 0.12 muons per initial 8 GeV incident proton. In addition, we have discussed a chicane system for the removal of unwanted secondary particles from the target region and showed that such a system can successfully remove particles with momentum > 1 GeV/c. Finally, we examined the influence of was in the whole performance. Our analysis indicated that a gas-filled buncher and rotator system could be used instead of the initial vacuum case. With appropriate increases in rf gradient, μ/p similar to the baseline can be obtained.

### Acknowledgments

The authors are grateful to K. T McDonald, X. Ding, H. K. Sayed, R. B. Palmer, H. Kirk and M. Palmer for many useful discussions. This work is supported by the U.S. Department of Energy, Contract no. DE-AC02-98CH10886.